\documentstyle[sprocl]{article}

\def\s0#1#2{\mbox{\small{$ \frac{#1}{#2} $}}}
\def\0#1#2{\frac{#1}{#2}}
\def\de{\delta}

\def\eq#1{(\ref{#1})}

\def\beq{\begin{equation}}
\def\eeq{\end{equation}}
\def\bea{\begin{eqnarray}}
\def\eea{\end{eqnarray}}

\begin{document}

\title{Deriving effective transport equations for non-Abelian plasmas}

\author{Daniel F. Litim \footnote{E-Mails: D.Litim@thphys.uni-heidelberg.de$,\ $Cristina.Manuel@cern.ch}${}^{,1}$ and Cristina Manuel ${}^{a,2}$}

\address{${}^1$Inst.\,f.\,Theoretische Physik, Philosophenweg 16, 
D-69120 Heidelberg, Germany.\\
${}^2$Theory Division, CERN, CH-1211 Geneva 23, Switzerland.}

\maketitle
\abstracts{Within classical transport theory, an approach is presented to derive systematically effective collision terms and noise sources for effective kinetic equations of non-Abelian plasmas. This procedure amounts to the `integrating-out' of fluctuations, and is applicable for in- and out-of-equilibrium situations. When applied to a hot non-Abelian plasma close to equilibrium we recover the collision integral and the noise source first obtained by B\"odeker. It is also shown that our approach, and hence B\"odeker's effective theory, is consistent with the fluctuation-dissipation theorem.
\vspace*{-8cm}
\begin{flushright}
{\normalsize HD-THEP-00-23\\ CERN-TH 2000/098}
\end{flushright}
\vspace*{7cm}
}

\section{Introduction}
Understanding the physics of a quark-gluon plasma at high temperature or density -- which is expected to be detected in the up-coming heavy ion experiments at RHIC and LHC -- requires reliable theoretical tools to describe transport phenomena in non-Abelian plasmas in- or out-of-equilibrium. Such methods are also of relevance for the physics of the early universe, if the baryon asymmetry could finally be understood within an electroweak framework.

In this talk\footnote{Invited talk given by DFL at 5th workshop on QCD, Villefranche-sur-Mer, 3-7 Jan 2000.} an approach is discussed which allows for a systematic derivation of effective kinetic equations for non-Abelian plasmas.\cite{LM}${}^{\mbox{-}}\,$\cite{FDT} The interest in developing a formalism based on a classical transport theory for non-Abelian gauge fields \cite{Heinz} is that main properties of a hot {\it quantum} plasma can already be understood within simple {\it classical} terms. Indeed, the soft non-Abelian gauge fields -- those having a huge occupation number -- can be treated as classical {\it fields}, while the hard gauge field modes can be treated as classical {\it particles}. This method, based on `integrating-out' fluctuations about some mean fields, has been known since long for Abelian plasmas \cite{K}, but a consistent extension to the non-Abelian case was longtime missing.

We apply our formalism to a hot plasma close to equilibrium. While the HTL effective theory is recovered in the Vlasov approximation \cite{KLLM}, the inclusion of leading order corrections due to fluctuations results in B\"odeker's effective theory.\cite{DB} For a more extensive discussion of this and related effective theories we refer to the discussion in ref.~\cite{DBP}. Some further approaches to study quark-gluon dynamics are considered in ref.~\cite{HeinzP}.

\section{Microscopic transport equations}
The starting point for a classical transport theory is to consider an ensemble of classical point particles carrying a non-Abelian charge $Q_a$, where the colour index runs from $a=1$ to $N^2-1$ for a SU($N$) gauge group. These particles interact self-consistently amongst each others, that is, through the classical gauge fields created by the particles. Their classical equations of motion are known as the Wong equations \cite{Wong}, 
\beq
m\0{d{\hat x}^\mu}{d\tau}={\hat p}^\mu \ ,\quad
m\0{d{\hat p}^\mu}{d\tau}=g {\hat Q}^a F_a^{\mu\nu} {\hat p}_\nu \ ,\quad
m\0{d{\hat Q}^a}{d\tau}=-g f^{abc} {\hat p}^\mu A^{b}_{\mu}{\hat Q}^c \
.\label{dQ}
\eeq
Here, $A_\mu$ denotes the gauge field, $F^{a}_{\mu\nu}=\partial_\mu A^{a}_\mu-\partial_\nu A^{a}_\mu+ gf^{abc}A^{b}_\mu A^{c}_\nu$ the corresponding field strength and $\hat z\equiv (\hat x,\hat p,\hat Q)(\tau)$ the world line of a particle. 
An ensemble of such particles can be described by their one-particle distribution function $f\sim\sum_i \int d\tau \delta(z-\hat z_i)$. Making use of (\ref{dQ}), its kinetic equation is \cite{Heinz}
\beq
\label{NA-f}
p^\mu\left(\0{\partial}{\partial x^\mu}
- g f^{abc}A^{b}_\mu Q^c\0{\partial}{\partial Q^a}
-gQ_aF^{a}_{\mu\nu}\0{\partial}{\partial p_\nu}\right) f(x,p,Q)=0 \ . 
\eeq
This Boltzmann equation is collisionless, since $df/d\tau=0$ (Liouville's theorem). However, it contains {\it effectively} collisions due to the long range interactions as shall become clear in the sequel. Eq.~(\ref{NA-f}) is completed by the Yang-Mills equation $D_\mu F^{\mu\nu}=J^\nu$, with the current $J[f]$ obtained self-consistently from the particles. In this classical picture, $f$ is a deterministic quantity once all initial conditions have been fixed. This is, however, not possible, and $f$ has to be considered as a {\it stochastic} (fluctuating) quantity instead. 

\section{Coarse graining}
Of particular interest for most physical applications are the {\it macroscopic} properties of such a plasma of particles. In this case, a more appropriate quantity to study would be a {\it coarse-grained} distribution function $\bar f$, whose dynamical equation is expected to be of the Boltzmann-Langevin type, 
\beq
\label{NA-f-cg}
p^\mu\left(\bar D_\mu-gQ_a\bar F^{a}_{\mu\nu}{\partial_p^\nu}\right) \bar f
=C[\bar f]+\zeta \ . 
\eeq
(We introduced the shorthands $D_\mu[A]f \equiv [\partial_\mu-g f^{abc}Q_c A_{\mu,b}{\partial ^Q_a}]f$, $\partial_\mu\equiv{\partial }/{\partial  x^\mu}$, $\partial^p_\mu\equiv{\partial }/{\partial  p^\mu}$ and $\partial ^Q_a\equiv{\partial }/{\partial  Q^a}$). A coarse-grained distribution function obtains from the microscopic one after a suitable smearing-out over some characteristic volume, which, when applied to (\ref{NA-f}), results in a collision integral $C[f]$ and a related source for stochastic noise $\zeta$. Physically speaking, these terms are expected because the coarse-graining integrates-out the modes within 
a coarse-graining volume. The interactions of such modes can result into additional effective interactions for the remaining long range modes. Also, the stochastic fluctuations of $f$ are smeared-out, resulting in a noise term $\zeta$. 

\section{Effective transport equations}
An effective kinetic equation, like (\ref{NA-f-cg}), can be derived systematically within classical transport theory, starting with (\ref{NA-f}). To that end, we perform an ensemble average in phase space which allows to split $f$ and the gauge fields into their mean field part and a fluctuating part
\bea
f(x,p,Q)&=& {\bar f}(x,p,Q) + \de f(x,p,Q) \ ,\\
A^\mu(x)&=& {\bar A}^\mu(x) + a^\mu(x) \ ,\label{NA-delta-a}
\eea
where ${\bar f} = \langle f \rangle$ and ${\bar A} = \langle A \rangle$. The mean value of the statistical fluctuations vanish  by definition,  $\langle \de f \rangle =0$ and $\langle a \rangle=0$. This separation into mean fields and statistical, random  fluctuations corresponds effectively to a split into long/short wavelength modes associated to the mean fields/fluctuations. Performing the ensemble average of (\ref{NA-f}) and the Yang-Mills equations gives \cite{LM,LM2}
\begin{eqnarray}
p^\mu\left(\bar D_\mu- gQ_a \bar F^a_{\mu\nu} 
 \partial _p^\nu\right) f&=&\left\langle\eta\right\rangle
+ \left\langle\xi\right\rangle \ . \label{NA-1}\\
\bar D_\mu \bar F^{\mu\nu}  + \left\langle J_{\mbox{\tiny fluc}}^{\nu}
\right\rangle &=&\bar J^\nu \ .\label{NAJ-1} 
\end{eqnarray}
In \eq{NA-1} and \eq{NAJ-1}, we collected all terms quadratic or cubic in the
fluctuations into 
\begin{eqnarray}
\eta &\equiv & g Q_a\, p^\mu 
\left[ (\bar D_\mu a_\nu- \bar D_\nu a_\mu)^a+ g f^{abc}a^b_\mu
a^c_\nu\right]\, \partial _p^\nu  \delta f(x,p,Q)  \ , \label{NA-eta}\\
\xi &\equiv & gp^\mu f^{abc}Q^c\left[\partial ^Q_a a_\mu^b\ 
\delta f(x,p,Q)\ + g a_\mu^a a_\nu^b\partial^\nu_p\bar f(x,p,Q)\right]
,\label{NA-xi}\\
J_{\mbox{\tiny fluc}}^{a,\nu}&\equiv & g \left[f^{dbc} 
\bar D^{\mu}_{ad}  a_{b,\mu} 
a_c^\nu  + f^{abc}  a_{b,\mu}\, \left( (\bar D^\mu a^\nu-\bar D^\nu
a^\mu)_c+ g f^{cde}a^\mu_d a^\nu_e \right) \right] .\quad \label{NA-Jfluc}
\end{eqnarray}
In addition, we need the dynamical equations for the fluctuations. They are derived from subtracting (\ref{NA-f}) from (\ref{NA-1}) as 
\begin{eqnarray}
p^\mu\left(\bar D_\mu - gQ_a\bar F^a_{\mu\nu}\partial _p^\nu\right)\delta f
&=&
g Q_a(\bar D_\mu a_\nu-\bar D_\nu a_\mu)^a p^\mu \partial ^p_\nu\bar f
\nonumber \\ && 
+ gp^\mu a_{b,\mu} f^{abc} Q_c\partial ^Q_a \bar f 
+\eta  + \xi - \left\langle\eta+\xi\right\rangle\label{NA-2}\\
\left(\bar D^2 a^\mu-\bar D^\mu(\bar D_\nu a^\nu)\right)^a
&+& 2 gf^{abc}
\bar F_b^{\mu\nu}a_{c,\nu}+J_{{\mbox{\tiny fluc}}}^{a,\mu}-\left\langle
J_{{\mbox{\tiny fluc}}}^{a,\mu}\right\rangle
=\delta J^{a,\mu} \ .\label{NAJ-2}
\end{eqnarray}
Eqs.~(\ref{NA-2}) and (\ref{NAJ-2}) are the master equations for all higher order correlation functions of fluctuations. (This hierarchy of dynamical equations is similar to the BBGKY hierarchy.) All the equal-time correlators of $\delta f$ can be derived from the basic definition of the Gibbs ensemble average in phase space.\cite{LM2} The most important correlator is the quadratic one $\langle\delta f\delta f\rangle_{t=t'}$ which can be expressed in terms of $\bar f$ and a two particle correlation function.\cite{LM2,LL} 

Some few comments. ($i$) The set of dynamical eqs.~(\ref{NA-1}) - (\ref{NAJ-2}) is equivalent to the original set of microscopic equations. It is exact in that no approximations have been performed so far, and can be applied to both in- or out-of-equilibrium situations. ($ii$) The original equations as well as their split into mean gauge fields and fluctuations have been shown to be consistent with the non-Abelian mean field gauge symmetry.\cite{LM2,KLLM} ($iii$) In order to obtain effective equations for the mean fields only, one has to integrate-out the fluctuations. This amount to solving explicitly the dynamical equations of $\delta f$ and the gauge field fluctuations for given initial conditions and to express the correlator terms explicitly as functions of the mean quantities, making use of the basic equal-time correlators. In this light, our procedure can be seen as a {\it derivation} of collision integrals. ($iv$) The dynamical equations for the fluctuations are non-linear as they involve fluctuations up to cubic order. Solving them will require some approximations. For small gauge coupling an expansion in powers of $g$ can be performed. If the fluctuations remain small within a coarse-graining volume, the second moment approximation can be used, which amount ultimately to a linearisation of the dynamics of the fluctuations.\cite{K} These types of approximations are systematic and consistent with the non-Abelian gauge symmetry of the mean fields. ($v$) In the Abelian limit $\langle\xi\rangle$ vanishes identically, while $\langle\eta\rangle$  reduces to the well-known Balescu-Lennard collision integral.\cite{K} ($vi$) In the limit where fluctuations are fully neglected, the set of equations are known as the (non-Abelian) Vlasov equations.\cite{Heinz,KLLM} 

\section{The hot non-Abelian plasma close to equilibrium}
We now turn to the specific example of a hot thermal plasma close to equilibrium. `Hot' implies that particle masses can be neglected to leading order $m(T)\ll T$, and that the gauge coupling, as a function of temperature, is very small $g(T)\ll 1$. The non-Abelian plasma is characterised by two dimensionless parameters, the gauge coupling $g$ and the plasma parameter $\epsilon$, which obtains as the ratio of the volume per particle over a Debye volume. The Debye (or screening) length describes the polarisation of the plasma. A small plasma parameter is mandatory for a kinetic description to be viable. Physically, this means that a large number of particles interact coherently within a Debye volume (quasi-particle behaviour). For a classical plasma, $\epsilon$ is an independent parameter related to the mean particle number and the gauge coupling.\cite{LM2} For a quantum plasma, the mean particle number can not be fixed arbitrarily, and $\epsilon$ scales as $\epsilon\sim g^3\ll 1$. This explains why a kinetic description, for quantum plasmas, is intimately linked to the weak coupling limit. 

To simplify our analysis we shall perform some approximations, all of which can be understood as a systematic expansion in $g$. After ensemble averaging, the distribution function $f$ is effectively coarse-grained over a Debye volume. Fluctuations of $\bar f$ within a Debye volume are parametrically suppressed by powers of $g$. We expand $f$ about the equilibrium distribution function $\bar f^{\rm eq}$ to leading order in $g$ as
\beq\label{Ansatz}
f=\bar f^{\rm eq}+g\bar f^{(1)}+\delta f\ .
\eeq
Solving (\ref{NA-1}) using (\ref{Ansatz}) for $\delta f= 0$ reproduces the HTL effective theory.\cite{KLLM} Two further approximations are employed beyond the HTL level, when $\delta f\neq 0$. We shall discard cubic correlator terms in (\ref{NA-1}), which are additionally suppressed by a power in $g$, in favour of quadratic ones. This corresponds to neglecting three-particle collisions in favour of binary ones. We also employ the second moment approximation -- setting $\eta=\langle\eta\rangle$ and $\xi=\langle\xi\rangle$ in (\ref{NA-2}) -- which means that the effects of collisions are neglected for the dynamics of the fluctuations. All these approximations can be systematically improved to higher order.
  
Let us consider colour excitations, described by the colour current density
\beq\label{Jdens}
{\cal J}^\mu_a(x,{\bf v})=g\, v^\mu\,\int dQ\,dp_0\,d|{\bf p}|\,{\bf p}^2\, 
\Theta(p_0)\,\delta(p^2)\, Q_a\, f(x,p,Q)\ ,
\eeq
and $v^\mu\equiv p^\mu/p_0=(1,{\bf v}), {\bf v}^2=1$. The current $J$ is obtained after an angle average over the directions of ${\bf v}$ as $J(x)=\int\frac{d\Omega_{\bf v}}{4\pi}{\cal J}(x,{\bf v})$. The colour measure $dQ$ is normalised $\int dQ=1$ and contains the Casimir constraints of the gauge group, such as $\int dQQ_aQ_b=C_2\delta_{ab}$ with $C_2$ the quadratic group Casimir (for more details, see ref.~\cite{LM2,KLLM}). The approximate dynamical equations for the fluctuations become
\begin{eqnarray}
&&\left(v^\mu \bar D_\mu \,  \delta   {\cal J}^\rho \right)_a \, = 
-m^2_D v^\rho v^\mu 
\left(\bar D_\mu a_0-\bar D_0 a_\mu\right)_a   -g f_{abc}  v^\mu a_\mu ^b
\bar {\cal J}^{c,\rho} 
\ ,\label{vD-dJa} \\
&&\left( \bar D^2 a^\mu-\bar D^\mu(\bar D a)\right)_a+2 g f_{abc} 
\bar F_b^{\mu\nu}a_{c,\nu}=\delta J_a^\mu  \ ,\label{dJa}
\end{eqnarray}
where the Debye mass $m_D$ obtains from the equilibrium distribution function and the group representation of the particles. Solving for the fluctuations in the present approximations yields a closed expression for $\delta {\cal J}$, and an iterative expansion in powers of the background fields for $a$.\cite{LM,LM2}  The seeked-for dynamical equation for the mean quasi-particle colour excitations is
\begin{eqnarray} \label{NAV-soft-mean}
v^\mu \bar D_\mu \bar {\cal J}^0+ m^2_D v^\mu \bar F_{\mu0}
&=& C_{\rm lin}[\bar {\cal J}^0] +\zeta(x,{\bf v})\ ,\label{NAV-curr}
\end{eqnarray}
where the linearised collision integral $C_{\rm lin}$ can now be evaluated explicitly, using the expressions for $a$ and $\delta {\cal J}$. The Yang-Mills equation remains unchanged at this order. For the collision integral, one finally obtains to linear order in the mean current, and to leading logarithmic order (LLO) in $(\ln 1/g)^{-1}\ll 1$
\begin{eqnarray}\nonumber
C_{\rm lin}[\bar {\cal J}_a^0](x,{\bf v})&=&
\left.gf_{abc}\left\langle a^b_\mu(x)\,
\delta {\cal J}^{c,\mu}(x,{\bf v})\right\rangle
\right|_{\bar A=0,\ {\rm linear \, in\,}\bar{J},\ {\rm LLO}}\\ \label{Clin}
&=&
-\gamma\int\frac{d\Omega'}{4\pi}I({\bf v},{\bf v}')\bar{\cal J}_a^0(x,{\bf v}')
\end{eqnarray}
with the kernel $I({\bf v},{\bf v}')=\delta^2 ({\bf v}-{\bf v}')-\frac{4}{\pi}({\bf v.v'})^2/\sqrt{1-({\bf v.v'})^2}$. Notice that the collision integral is local in coordinate space, but non-local in the angle variables. The rate $\gamma=\frac{g^2T}{4\pi}\ln 1/g$ is (twice) the hard gluon damping rate. In obtaining (\ref{Clin}), we introduced an infra-red cut-off $\sim g m_D$ for the elsewise unscreened magnetic sector. This result has been first obtained by B\"odeker in ref.~\cite{DB}. In addition, the source for stochastic noise $\zeta$ in (\ref{vD-dJa}) can be identified as
\bea\label{zeta}
\zeta_a(x,{\bf v})&=&
\left. gf_{abc} a^b_\mu(x)\,
\delta {\cal J}^{c,\mu}(x,{\bf v})\right|_{\bar A=0,\,\bar J=0}\\
\label{noise1}
\langle \zeta^a(x,{\bf v})\zeta^b(y,{\bf v}')\rangle &=&
2\gamma \,m_D^2 \,T \,I({\bf v},{\bf v}')\,
\delta^{ab}\,\delta (x-y)\ ,
\eea
and the self-correlator (\ref{noise1}) has been evaluated from (\ref{zeta}) to LLO. Notice that its strength is determined by the kernel of the dissipative process. Next, we show the close relationship to the fluctuation-dissipation theorem (FDT). 
\section{Fluctuation-dissipation theorem}
It is well-known that dissipation in a (quasi-stationary) plasma is intimately linked to the fluctuations, a link which is given by the FDT. For the coarse-grained kinetic equation (\ref{NA-f-cg}) one may ask how the spectral functions of $\zeta$ and $f$ have to be related to $C[f]$ in order to satisfy the FDT. This question has been considered in ref.~\cite{FDT}. Within classical transport theory, the FDT is implemented in a straightforward way.\cite{LL} The pivotal element is the kinetic entropy $S[f]$. We consider small deviations of $f$ from the equilibrium, $f=f^{\rm eq}+\Delta f$. The entropy, stationary at equilibrium, is expanded to quadratic order in $\Delta f$, $S=S_{\rm eq}+\Delta S$. We define the thermodynamical force as $F(z)=-\delta (\Delta S)/\delta (\Delta f(z))$, where $z\equiv (x,{\bf p}, Q)$. Given $C[f](z)$, it can be shown \cite{FDT,LL} that 
\beq\label{noisenoise}
\langle\zeta(z)\zeta(z')\rangle=-\left(\frac{\delta C(z)}{\delta F(z')}+\frac{\delta C(z')}{\delta F(z)}\right)\ 
\eeq
is the  noise self-correlator for $\zeta$ compatible with the FDT. For the particular example studied above, (\ref{noise1}) follows from inserting (\ref{Clin}) into (\ref{noisenoise}). This proves that B\"odeker's effective theory is compatible with the FDT. Furthermore, the correlator $\langle\Delta f\Delta f\rangle|_{t=t'}$ can be derived along the same lines and agrees, to leading order, with $\langle\delta f\delta f\rangle|_{t=t'}$ from the Gibbs ensemble average. This guarantees that our formalism is consistent with FDT.

\section{Discussion}
We have presented a systematic approach to derive effective transport equations for non-Abelian plasmas. The procedure amounts to the `integrating-out' of fluctuations about some mean fields. Most importantly, the formalism is applicable for both in- and out-of-equilibrium situations, simply because the main statistical information as encoded in the equal-time correlators of fluctuations does not depend on the mean distribution function being thermal, or not. At the same time, the formalism -- and certain approximations to it -- respects the underlying non-Abelian mean field symmetry, which is at the basis for any reliable computation. 

For the close-to-equilibrium plasma, we have shown how B\"odekers effective theory emerges to leading logarithmic accuracy. The collision integral as well as the necessary noise source have been derived explicitely from the microscopic theory. In addition, it is established on general grounds that the formalism (and hence B\"odeker's effective theory) is consistent with the fluctuation-dissipation theorem.

As a final comment we point out that the effective theory is the same for a classical or a semi-classical/quantum plasma, differing only in the equilibrium distribution function (Maxwell-Boltzmann vs.~Bose-Einstein or Fermi-Dirac), and hence in the corresponding value for the Debye mass. The sole `quantum' effect which entered our computation resides in the non-classical statistics of the particles, which is all that is needed to correctly describe a hot {\it quantum} non-Abelian plasma close to equilibrium at the present order of accuracy.


\section*{References}
\begin{thebibliography}{99}

\bibitem{LM} 
D.F.~Litim and C. Manuel, 
    Phys.~Rev.~Lett.~{\bf 82} (1999)  4981.

\bibitem{LM2} 
D.F.~Litim and C. Manuel, 
    Nucl.~Phys.~{\bf B 562} (1999) 237.

\bibitem{FDT} 
D.F.~Litim and C. Manuel, 
    Phys.~Rev.~{\bf D} (to appear), {hep-ph/9910348}.

\bibitem{Heinz}
 U.~Heinz, Phys. Rev. Lett. {\bf 51} (1983) 351; Ann. Phys. 
{\bf 161} (1985) 48.

\bibitem{K}
Yu. L. Klimontovich, ``{\it Statistical Physics}", Harwood  Publ, (1986).

\bibitem{KLLM}
P.R. Kelly, Q. Liu, C. Lucchesi and C. Manuel, 
Phys. Rev. Lett {\bf 72} (1994) 3461; Phys. Rev. {\bf D 50} (1994) 4209.

\bibitem{DB}
D.~B\"odeker, Phys. Lett. {\bf B426} (1998) 351.

\bibitem{DBP} D.~B\"odeker, these proceedings.

\bibitem{HeinzP} U.~Heinz, these proceedings.

\bibitem{Wong}
S. Wong, Nuovo Cim.~{\bf 65A} (1970) 689.

\bibitem{LL} L.~Landau and E.M.~Lifshitz, ``{\it Statistical Physics part 1}", Oxford (1981).

\end{thebibliography}
\end{document}